# Combining Modified Weibull Distribution Models for Power System Reliability Forecast

Ming Dong, *Senior Member, IEEE*, and Alexandre B. Nassif, *Senior Member, IEEE*

***Abstract*—Under the deregulated environment, electric utility companies have been encouraged to ensure maximum system reliability through the employment of cost-effective long-term asset management strategies. Previously, the age based Weibull distribution has been used vastly for modeling and forecasting aging failures. However, this model is only based on asset age and does not consider additional information such as asset infant mortality period and equipment energization delay. Some works on modifying Weibull distribution functions to model bathtub-shaped failure rate function can be practically difficult due to model complexity and inexplicit parameters. To improve the existing methods, this paper proposes four modified Weibull distribution models with straightforward physical meanings specific to power system applications. Furthermore, this paper proposes a novel method to effectively evaluate different Weibull distribution models and select the suitable model(s). More importantly, if more than one suitable model exists, these models can be mathematically combined as a joint forecast model, which could provide better accuracy to forecast future asset reliability. Finally, the proposed approach was applied to a Canadian utility company for the reliability forecast of electromechanical relays and distribution poles to demonstrate its practicality and usefulness.**

## I. Introduction

ELECTRIC utility companies have had to adapt to the deregulated environment and find ways to reduce overall cost while maintaining system reliability performance. To achieve this goal, understanding and forecasting reliability trends of different asset populations is the key. Sophisticated and optimal asset management measures can only be established based on the accurate forecasting of asset reliability change in the future.

Previously, the asset age based Weibull distribution has been the traditional statistical tool to model equipment aging failures in reliability engineering [1]–[4]. However, this classic model cannot effectively incorporate additional information such as asset health condition data, asset warranty, energization delay, asset infant mortality period and minimum spare requirements which many electric utility companies often need to consider for power system asset management. Many recent works to improve Weibull Distribution models [6]–[10] were focused on modeling bathtub-shaped failure rate function. Compared to a standard Weibull function, the bathtub-shaped function can describe the initial asset infant mortality period and the stable low failure rate period before entering into the wear-out period that a standard two-parameter Weibull distribution is able to depict. To achieve this, these previous works proposed modified probability density functions (PDF) that are much more complicated than traditional Weibull PDF [9]. Due to this complication, the estimation of these parameters can be sometimes difficult, computationally costly and inaccurate due to potential over-fitting with multiple parameters. Lack of explicit physical meanings of these parameters made its application even more difficult to electric utility engineers.

To overcome the above problems, this paper proposes four modified Weibull distribution models, which are based on the standard two-parameter Weibull distribution model, except that they can include asset condition information and two additional shifting parameters. These modifications are mathematically simple and straightforward. More importantly, the two shifting parameters have clear physical meanings and can be estimated by domain experts such as utility equipment or maintenance engineers. These parameters reflect the average failure rate during the initial infant mortality period and the stable low failure rate period respectively. In a way, the proposed method represents the bath-tub properties using a different but much simpler approach. Furthermore, these individual modified models can be combined to increase the forecasting flexibility and robustness. The main contributions of this paper are:

- Four modified Weibull distribution models with straightforward physical meanings that incorporate additional information and considerations specific to power system applications;
- A novel method to measure the forecast accuracy of different Weibull distribution models. Based on this method, suitable Weibull distribution model(s) can be identified for a specific asset population;
- A novel method to combine different Weibull distribution models as a joint forecast model.

The proposed approach is described as follows and its flowchart is shown in Fig. 1. Firstly, the operation status data of a specific asset population is analyzed and converted to a cumulative failure probability table. This table is then split into training pairs and testing pairs. Training pairs are used to model

Manuscript received June 3, 2018; revised September 11, 2018; accepted October 19, 2018. Date of publication October 24, 2018; date of current version February 18, 2019. Paper no TPWRS-00846-2018. *(Corresponding author: Ming Dong.)*

M. Dong is with the ENMAX Power Corporation, Calgary, AB T2G 4S7, Canada (e-mail: mingdong@ieee.org).

A. B. Nassif is with the ATCO Electric, Edmonton, AB T5J 2V6, Canada (e-mail: nassif@ieee.org).

Color versions of one or more of the figures in this paper are available online at http://ieeexplore.ieee.org.

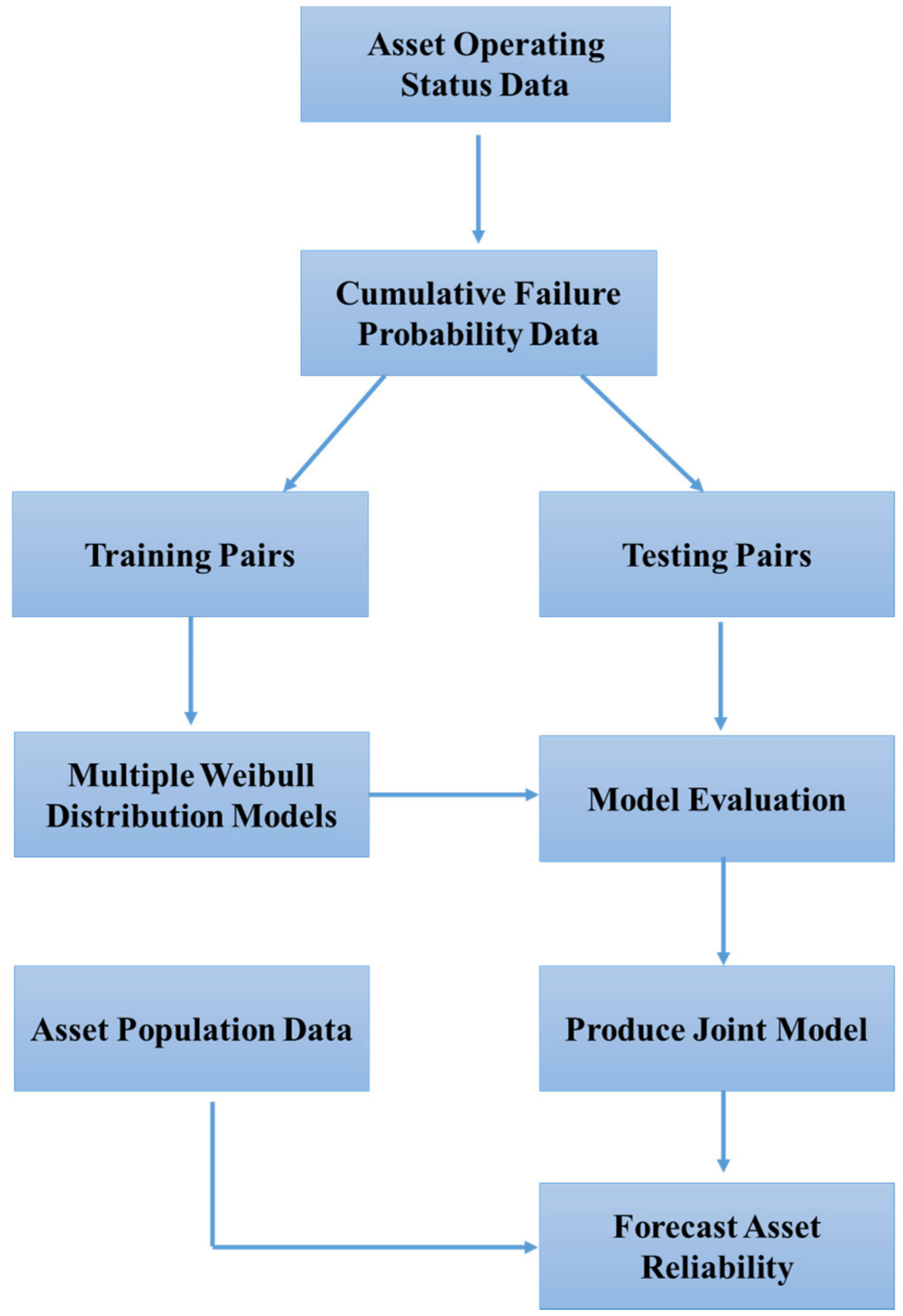

Fig. 1. Flowchart of the proposed approach for power system asset reliability forecast.

the asset failure progression using the proposed Weibull distributions models. Testing pairs are used to evaluate the forecast accuracy of the distribution models. The suitable Weibull distribution model(s) can be identified based on the model evaluation results. When there are more than one suitable Weibull distribution models, the models can be mathematically combined as a joint model. In the next stage, the combined joint model is applied to the broader asset population data and the future reliability change of this population is forecasted.

This paper first describes the required asset operation status dataset as the foundation of this work. It then reviews the standard Weibull distribution for aging analysis. Based on this review, four modified Weibull distribution functions (health index based, X-shifting, Y-shifting and XY-shifting) are proposed and discussed considering practical power system conditions. Subsequently, it reviews the existing methods to calculate two-parameter Weibull distribution parameters, discuss the process of creating training and testing pairs and explain the estimation of the two shifting parameters. After this, the proposed novel methods of evaluating and combining Weibull distribution models are presented along with the process of forecasting asset reliability by using the produced joint forecast model. Finally, this paper provides two examples of applying proposed methods to forecast future electromechanical relay failures and distribution pole failures, as well as planning spare requirements and replacement strategies for a Canadian utility company.

## II. Asset Operation Status Dataset

In recent years, to exploit the value embedded in data, many electric utility companies have employed sophisticated Computerized Maintenance Management Systems (CMMS) to track and store various asset data [11], [12]. In particular, asset reliability forecast requires the asset operation status data categorized by asset type and population. Asset operation status data should include age and operating statuses (i.e., working or failed). The data is continuously recorded as new assets get installed and old assets retire due to failures. Similar to any data mining algorithms, the quality of input data will significantly affect the accuracy of data observations and analysis [13]. Understanding the requirements of asset operation status data and preparing qualified asset operation status datasets is a key step towards the successful implementation of the methods proposed in this paper. It should be noted that:

- One type of asset often contains many sub-types due to different technology adoptions and manufacturing standards. The sub types should be manually identified by utility asset engineers based on equipment domain knowledge and separated into different datasets. For example, underground cables are mainly paper insulated cables, cross-linked Polyethylene (XLPE) or Ethylene Propylene Rubber (EPR) [14]. Commonly, utilities operate these different types of cables simultaneously as they were installed at different times in the past. These different types of cables display different failure characteristics and should be viewed as different asset types and analyzed separately. A second example is mineral oil immersed power transformers. Power transformers manufactured before 1970 are rated 55 °C; power transformers manufactured throughout the 1970s are rated 55/65 °C; after that, they became rated 65 °C till today [15]. Since the failure characteristics of transformers are closely related to winding and top-oil temperature rises, power transformers should be separated into different datasets by temperature rating or approximately by manufacturing time.
- Installation methods affect asset lifecycle. For example, underground cables can be direct buried in soil or installed in PVC or metal conduit. In general, the direct buried cables are more prone to failure due to soil corrosion and moisture ingress [16]; distribution and transmission power poles are constructed as tangent structures (carrying straight conductors), angle structures (having deflection angles), dead-end structures (at the end of the line) or transformer poles (carrying mounted transformers). These structures incur different mechanical forces which lead to different failure characteristics. Therefore, assets with different installations should also be separated into different populations and form different operation status datasets.
- Utility companies may perform inspection or testing for certain asset types and populations. If available, the latest inspection/testing data can be linked with the age and operating status data. This is because the modified Weibull distribution models proposed in this paper can incorporate asset condition data and derive health index based models.

TABLE I
ASSET OPERATION STATUS DATASET FOR 138 kV IN-DUCT XLPE TRANSMISSION UNDERGROUND CABLE

| Cable ID | Age | Operating Status | Partial Discharge Result | Neutral Condition | Splice Condition |
|---|---|---|---|---|---|
| 1 | 10 | Working | Good | Good | Good |
| 2 | 11 | Working | Good | Medium | Good |
| 3 | 17 | Working | Medium | Good | Medium |
| 4 | 37 | Failed | Medium | Good | Good |
| 5 | 45 | Working | Good | Poor | Medium |
| 6 | 43 | Working | Good | Good | Medium |
| 7 | 52 | Failed | Poor | Poor | Poor |
| 8 | 25 | Failed | Good | Good | Poor |
| 9 | 35 | Working | Good | Poor | Good |
| 10 | 40 | Working | Good | Good | Good |
| … | … | … | … | … | … |

Currently in the North American power industry, for a specific asset type, health indexes are often developed using weighting factors for different conditional attributes [17]. As an example, for underground cables, partial discharge test results, neutral conditions and splice conditions can be ranked from poor to good and quantified to produce underground cable health indexes using weighted average. The determination of weighting factors should be based on the analysis from domain experts such as equipment and maintenance engineers. Discussions with the company's asset management/maintenance department, company's field crews, equipment manufacturers or peer utility companies are often required to determine confident weighting factors.

- The asset operation status dataset used by the proposed methods does not have to contain the entire population for a specific type of asset. Oftentimes, a randomly sampled population fraction with hundreds of records will be sufficient to establish a confident Weibull distribution model. Data completeness is another consideration. Sometimes only limited members in a population have inspection or testing data and these members should be utilized for establishing the model. The established model can still be applied to the entire population at a later stage.

As an example, a 138 kV in-duct XLPE transmission underground cable operation status dataset is shown in Table I which follows the above discussed data requirements.

## III. STANDARD AGE BASED WEIBULL DISTRIBUTION MODEL

The Weibull distribution is a continuous probability distribution widely used in the field of reliability engineering [1–4]. The standard probability density function of equipment age is

$$f\left(A\right) = \left(\frac{\beta}{\alpha}\right)\left(\frac{A}{\alpha}\right)^{\beta-1} e^{-\left(\frac{A}{\alpha}\right)^{\beta}}, A \geq 0 \quad (1)$$

where $\alpha > 0$ is the scale parameter; $\beta > 0$ is the shape parameter; $A$ in this application is the equipment age; $f(A)$ is the probability of failure at age $A$. The corresponding cumulative Weibull distribution function is given as:

$$F\left(A\right) = 1 - e^{-\left(\frac{A}{\alpha}\right)^{\beta}}, \ A \geq 0 \quad (2)$$

where $F(A)$ is the cumulative probability of failure.

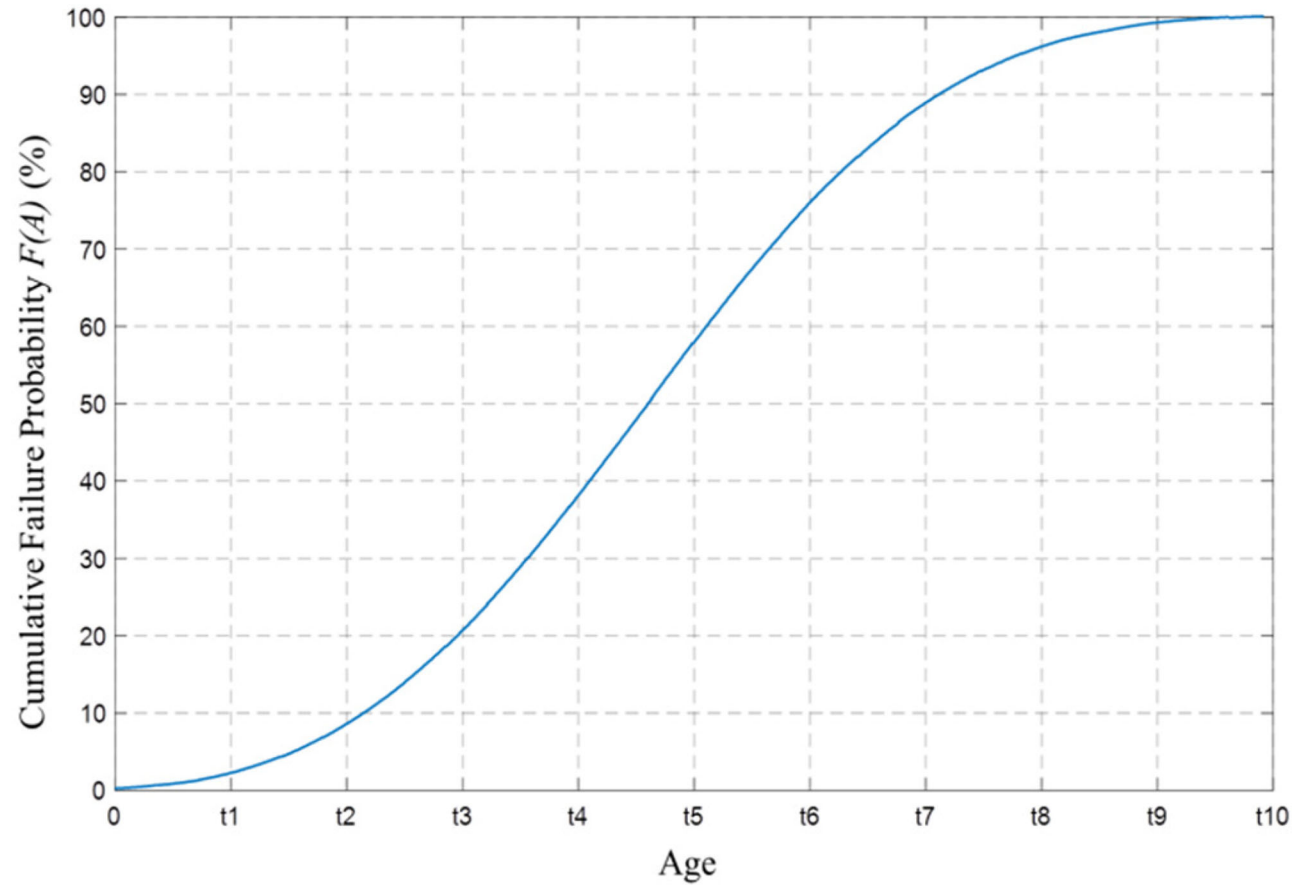

Fig. 2. Standard age based Weibull distribution.

TABLE II
AN EXAMPLE OF CUMULATIVE FAILURE PROBABILITY VERSUS AGE

| A | $\hat{F}$(A) |
|---|---|
| 0 | 0.0% |
| 1 | 0.0% |
| 2 | 0.8% |
| 3 | 1.5% |
| 4 | 1.5% |
| 5 | 2.2% |
| 6 | 3.5% |
| 7 | 5.7% |
| … | … |
| 32 | |
| 33 | 98.2% |
| 34 | 99.1% |
| 35 | 100% |

Fig. 2 shows an example of cumulative Weibull distribution function for a generic type of equipment. As age progresses, the cumulative probability of failure increases until it reaches 100% at a later age. This is a statistical description of an asset aging failure process and can be used for equipment survival analysis [18].

This model can be derived from the asset operation status dataset discussed in Section II. For a specific age $A$, the observed cumulative failure probability $\hat{F}(A)$ is

$$\hat{F}\left(A\right) = \frac{N_{fA}}{N_{fA} + N_{wA}} \times 100\% \quad (3)$$

where $N_{fA}$ is the total number of failed assets in the dataset with an age less than or equal to $A$ and $N_{wA}$ is the total number of working assets in the dataset with an age greater than age $A$. In other words, $N_{fA}$ represents the assets that failed before age $A$ and $N_{wA}$ represents the assets that survived past age $A$. Based on this calculation, a new table can be produced to depict the relationship of Age $A$ and the observed corresponding cumulative failure probability $\hat{F}(A)$.

Table II shows an example of cumulative failure probability vs. age. $\hat{F}(A)$ increases to 100% at the age of 35. It should be noted that sometimes, due to data availability, the observed $\hat{F}(A)$ may not immediately change between consecutive years such as when $A = 3$ and 4 in this example. This does not impact the modelling as the observed $\hat{F}(A)$ is based on frequency statistics and does not always comply with $F(A)$. Table II will be split into training and testing pairs. The training pairs will be fed into the Weibull distribution modeling module as shown in Fig. 1, following the methods explained in Section V to estimate the parameters $\alpha$ and $\beta$.

## IV. Modified Weibull Distribution Models

This paper proposes a few modified Weibull distribution models. The physical meanings of these distributions and their applications are discussed in this section.

### A. Two-Parameter Health Index Based Weibull Distribution

Equation (2) considers age as the sole factor in the asset's failure progression modeling. This could be accurate for assets that are used in a near homogenous environment where operating conditions such as loading level and temperatures are about the same. However, for a large power system where operating conditions vary significantly, it is not uncommon to find a young asset with higher failure probability and an old asset with a lower failure probability. In cases like this, the classic age based Weibull distribution modeling will be less capable to provide an accurate forecast. However, as mentioned in Section II, if additional asset condition information is available, the independent variable *A* can be converted to health index *H* as shown below.

$$\begin{cases} F\left(H\right) = 1 - e^{-\left(\frac{H}{\alpha}\right)^{\beta}}, & H \geq 0 \\ H = W_a A + \sum_{i=1}^{t} W_i C_i, & W_a + \sum_{i=1}^{t} W_i = 1 \end{cases} \tag{4}$$

where *A* is the asset age normalized to the same scale of conditional attributes $C_i$; $W_a$ is the weighting factor for asset age; $W_i$ is the weighting factor for $i_{th}$ condition attribute; $C_i$ is the $i_{th}$ normalized condition attribute. The above equation uses different weighting factors to combine age and other condition data to generate an index between 0 and 100 [17]. Similar to using age alone, the health index *H* progresses from 0 (brand new, completely healthy) to 100 (most unhealthy observation in the system). The higher *H* is, the higher the corresponding failure probability *F(H)* is. Therefore, the standard age based Weibull distribution (2) is just a special case of (4) when condition weighting factors $W_i = 0$ and $W_a = 1$. Fig. 3. shows an example of health index based Weibull distribution.

This model can be derived from the asset operation status dataset discussed in Section II. For a specific health index *H*, the observed cumulative failure probability

$$\hat{F}\left(H\right) = \frac{N_{fH}}{N_{fH} + N_{wH}} \times 100\% \tag{5}$$

where $N_{fH}$ is the total number of failed asset in the dataset with a health index less than or equal to *H*; $N_{wH}$ is the total number of working asset in the dataset with a health index greater than *H*. Based on this calculation, Table III can be produced to depict the relationship between health index *H* and cumulative failure probability *F(H)*. Table III shows an example of cumulative failure probability vs. health index.

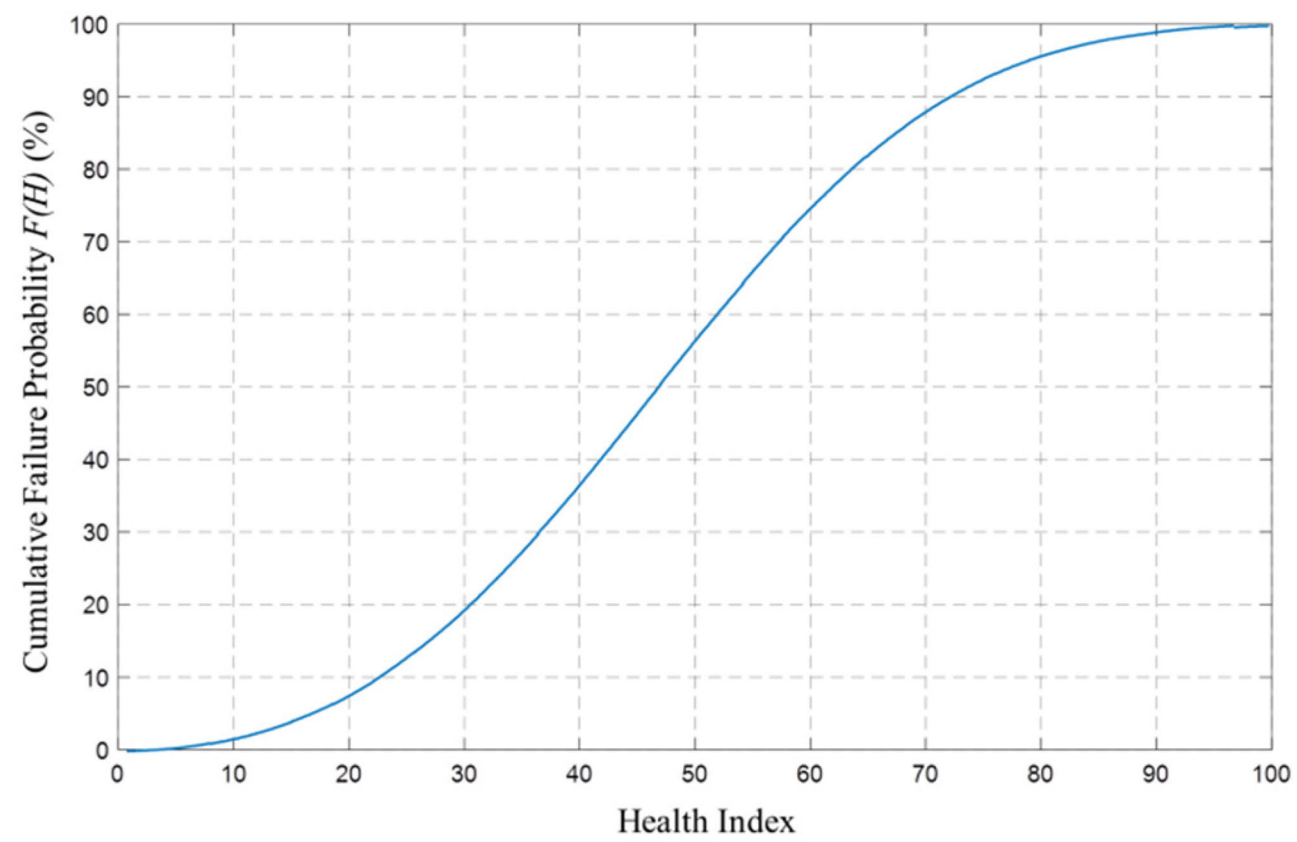

Fig. 3. Two-parameter health index-based Weibull distribution.

TABLE III
An Example of Cumulative Failure of Probability Versus Health Index

| H | F(H) |
|---|---|
| 0 | 0.0% |
| 1 | 0.3% |
| 2 | 2.6% |
| 3 | 2.9% |
| 4 | 6.5% |
| 5 | 9.3% |
| 6 | 9.5% |
| …. | … |
| 97 | 98.7% |
| 98 | 99.0% |
| 99 | 99.3% |
| 100 | 100% |

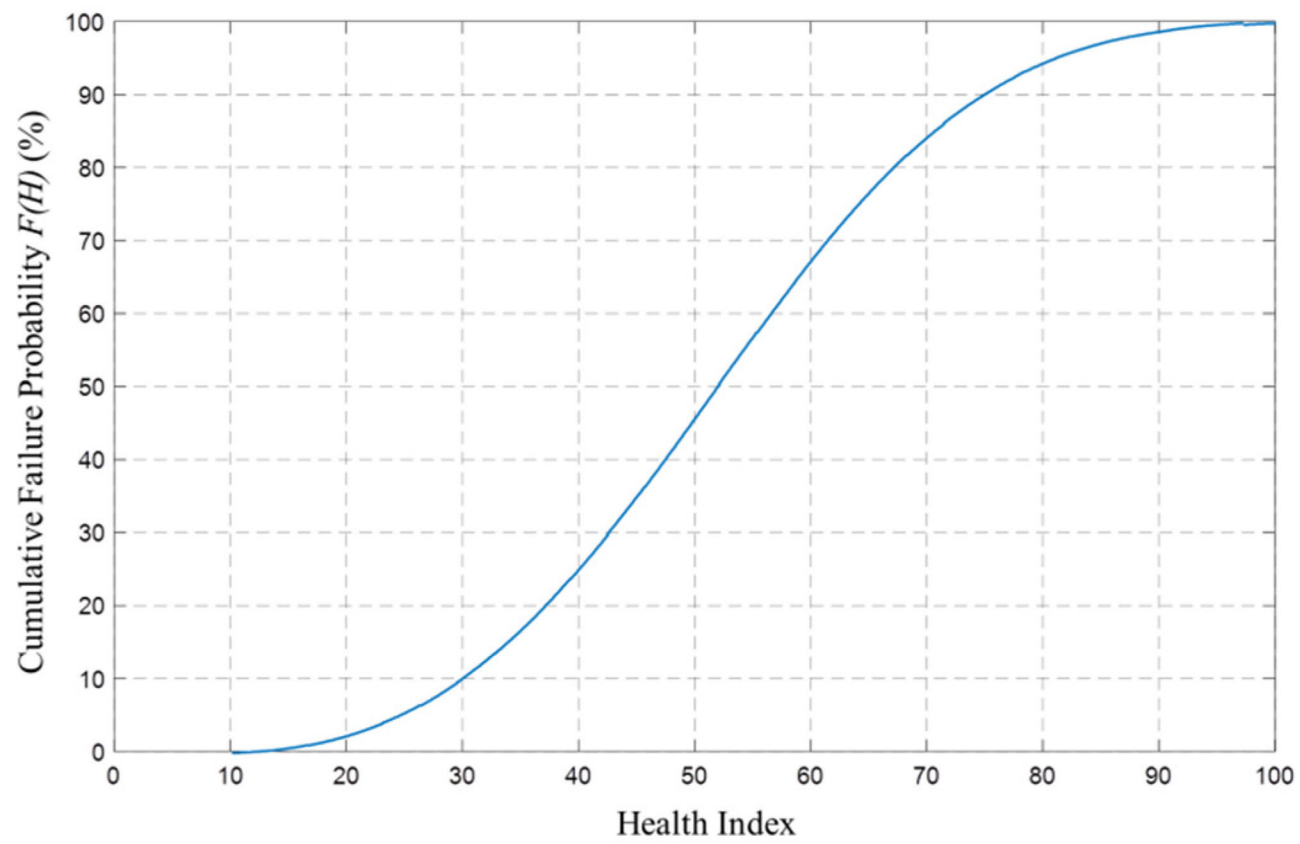

Fig. 4. X-Shift Weibull distribution.

### B. X-Shift Weibull Distribution

Compared to (4), X-Shift Weibull distribution has one additional parameter $\gamma$ $(\gamma \geq 0)$. This parameter $\gamma$ shifts the Weibull distribution curve along the *X* axis. Fig. 4. shows an example of X-Shift Weibull distribution. Mathematically, it is given as:

$$\begin{cases} F\left(H\right) = 1 - e^{-\left(\frac{H-\gamma}{\alpha}\right)^{\beta}}, & H \geq 0 \\ H = W_a A + \sum_{i=1}^{t} W_i C_i, & W_a + \sum_{i=1}^{t} W_i = 1 \end{cases} \tag{6}$$

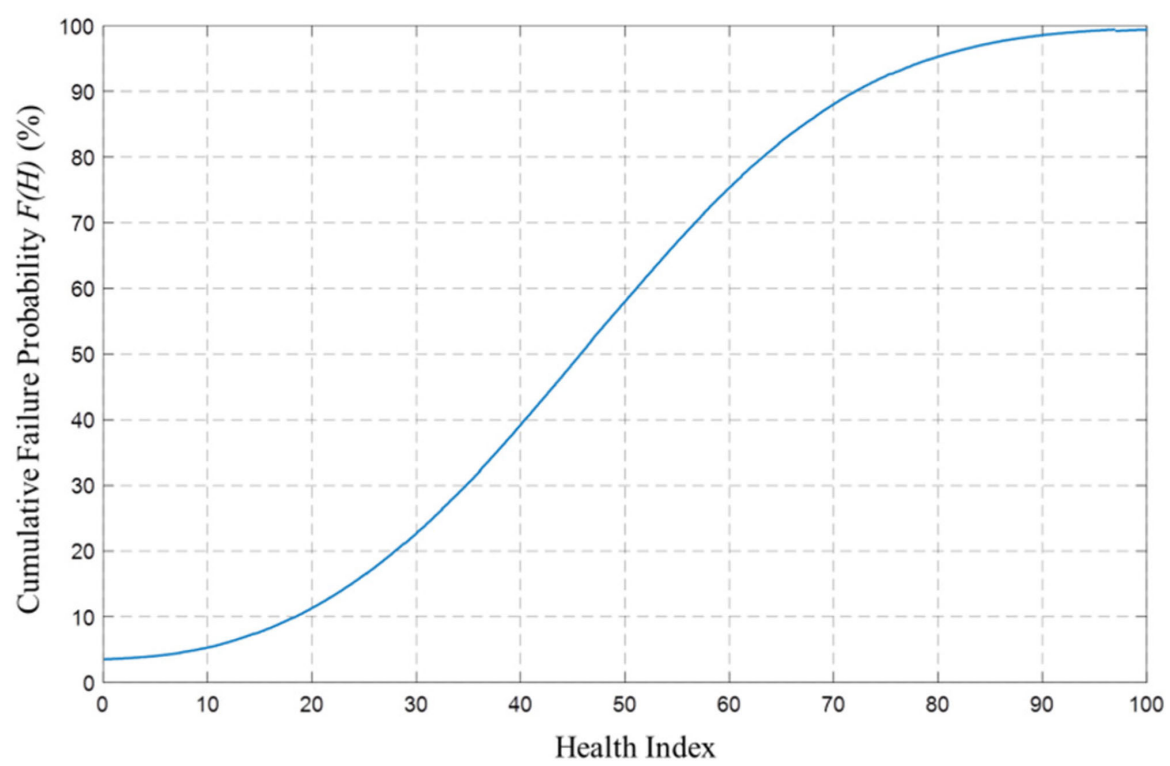

Fig. 5. Y-Shift Weibull distribution.

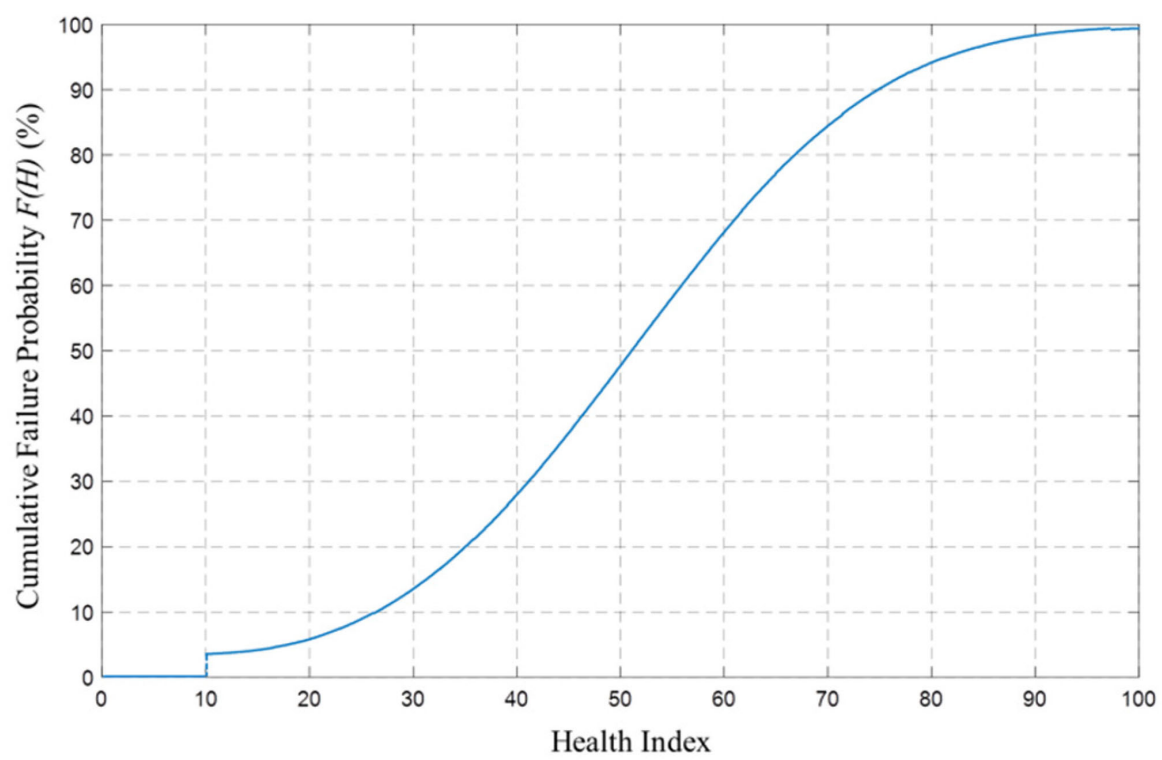

Fig. 6. XY-Shift Weibull distribution.

The physical meanings of parameter $\gamma$ in the power system context can be:

- A failure-free period: some assets naturally have an extremely low failure probability at a young age. This is more pronounced for assets in a protected, controlled and sometimes enclosed environment, for example, indoor GIS switchgears almost never fails in the first few years.
- Asset warranty and insurance: many assets have manufacturer warranty. From the utility perspective, this appears as the "failure-free" period. Even if there is a failure within the warranty period, the manufacturer will cover the material and replacement cost.
- Energization delay: after the installation of an asset, there could be an energization delay before this asset is energized and put into actual use. For example, an underground feeder system is often constructed in stages. The cable is often installed at an early stage and could stay in place for several months or even years before the other parts of system get constructed later. Only then, the system will be energized. This is common practice for many green field projects. In this case, cable will not fail before energization.

### C. Y-Shift Weibull Distribution

Compared to (4), Y-Shift Weibull distribution has one additional parameter $\delta$. This parameter $\delta$ shifts the Weibull Distribution curve along the *Y* axis. Fig. 5 shows an example of Y-Shift Weibull distribution. Mathematically, it is given as:

$$\begin{cases} F(H) = 1 + \delta - e^{-\left(\frac{H}{\alpha}\right)^{\beta}}, & H \geq 0,\ F(H) \leq 1 \\ H = W_a A + \sum_{i=1}^{t} W_i C_i, & W_a + \sum_{i=1}^{t} W_i = 1 \end{cases} \tag{7}$$

When $\delta < 0$, this distribution curve is shifted downward and the physical meaning becomes very similar to the X-shift Weibull distribution where a starting period of time has no positive failure probability (negative failure probability does not have a statistical meaning and can be capped to zero).

In reliability engineering, some assets could quickly fail at the initial period and the failure probability will reduce to a low and stable stage past this initial period, this period is called the infant mortality period [19], [20]. When $\delta > 0$, this distribution curve is shifted upward and the physical meaning is to capture the initial failure probability associated with the infant mortality period. Reflected by the cumulative distribution function, this initial failure can be described as a greater than zero cumulative failure probability $\delta$ when *H* is zero. Another occasion for applying a positive $\delta$ is when making spare forecast for critical assets for which a minimum spare level is always required to ensure maximum system reliability.

### D. XY-Shift Weibull Distribution

Compared to (4), XY-Shift Weibull distribution has two additional parameters $\gamma$ and $\delta$ to allow the Weibull distribution curve to shift along both *X* and *Y* axis. Fig. 6. shows an example of Y-Shift Weibull distribution. Mathematically, it is given as:

$$\begin{cases} F(H) = 1 + \delta - e^{-\left(\frac{H-\gamma}{\alpha}\right)^{\beta}} + \delta, & H \geq 0,\ F(H) \leq 1 \\ H = W_a A + \sum_{i=1}^{t} W_i C_i, & W_a + \sum_{i=1}^{t} W_i = 1 \end{cases} \tag{8}$$

XY-Shift Weibull distribution combines the characteristics of X-Shift Weibull distribution and Y-Shift Weibull distribution. For example, it can be used when an asset population has an energization delay as well as a failure probability in its infant mortality period.

## V. Estimation of Model Parameters

The cumulative failure probability data as shown in Tables II and III can be used to estimate Weibull distribution parameters and establish the models. However, not all data in the tables should be used to establish models. This is because this paper uses a unique method to evaluate and combine multiple Weibull distribution models. Similar to typical supervised machine learning processes [21], the cumulative failure probability data should be further split into two parts:

- *Training pairs:* the data pairs that are used to estimate Weibull distribution parameters. A certain percentage of the data, for example 80%, is randomly selected and grouped as training pairs. When selecting the training data, a uniform distribution is applied to sample the entire range of *H* and *F(H)*. This is to ensure the entire failure progression is modeled with enough supporting data.
- *Testing pairs:* the remaining data pairs that are used later to measure the accuracy of a model, as discussed in Section VI.

It should be noted statistically the more training pairs are used, the more accurate the models can become. Therefore, the training accuracy depends on the availability and completeness of training data [21]. This is the reason nowadays utility companies are encouraged to employ CMMS system to accurately track and classify failure history for different asset types so that enough training data can be established and utilized for parameter estimation and failure projection.

There are many existing ways of estimating a two-parameter Weibull distribution function's parameters $\alpha$ and $\beta$. Traditionally, Maximum Likelihood Estimation (MLE) and Least Square Estimation (LSE) were used in previous works [22]. In addition to MLE and LSE, there are other optimization algorithms for estimating Weibull distribution parameters such as simulated annealing algorithm [23] and genetic algorithms [24]. References [25] and [26] provided a detailed comparison of a few methods including MLE and LSE for Weibull distribution parameter estimation. Given many existing works on this topic, the selection and implementation of parameter estimation methods for two-parameter Weibull distribution functions are not discussed in detail in this paper. In fact, any parameter estimation method can work under the proposed framework for performance evaluation and joint forecast modeling as discussed next in Section VI.

Compared to the two-parameter Weibull distribution functions, the proposed X-Shift, Y-Shift and XY-Shift Weibull distribution functions have additional shifting parameters $\gamma$ or $\delta$. Not like $\alpha$ and $\beta$, they do not have to be mathematically estimated. This is because the shifting parameters are not random parameters and have clear physical meanings. Instead, the best way should be initializing them based on human expert's domain knowledge and then refining the values through evaluation later. Practically, a recommended way is to first estimate the approximate numerical ranges of $\gamma$ or $\delta$ based on past experience and prior knowledge in the utility company. Then $\gamma$ or $\delta$ can be discretized within these ranges. For example, in review of historical projects in a company, it is known that the typical energization delay for a transmission cable due to project execution is approximately 1–2 years. In this case, $\gamma$ can be initialized to 1.0, 1.25, 1.5, 1.75 and 2.0 with equal interval 0.25. For any of these five $\gamma$ values, there will be a corresponding X-Shift Weibull distribution model. Each model only has two unknown parameters $\alpha$ and $\beta$ and can therefore be estimated using the two-parameter estimate methods discussed above. The produced models can then be evaluated, selected and combined as one joint model, as discussed in Section VI. More examples of initializing $\gamma$ or $\delta$ can be found in Section VIII.

## VI. Evaluating and Combining Weibull Distribution Models

As discussed in Section V, the cumulative failure probability data is split to two parts: training pairs and testing pairs. This section explains how to use the testing pairs to evaluate the accuracy of the established models and further establish a joint model by combining the suitable models.

### A. Evaluating Weibull Distribution Models

Suppose $S$ is the set of established Weibull distribution models with different forms and parameters, given by:

$$S = \{F_1, F_2, F_3 \ldots\} \tag{9}$$

For each model $F_i$, the Mean Squared Error (MSE) for $n$ testing pairs is given by:

$$MSE = \frac{1}{n} \sum_{i=1}^{n} \left[\hat{F}(x_i) - F(x_i)\right]^2 \tag{10}$$

Only testing pairs are used for evaluation because testing pairs were not used previously for the model's parameter estimation and are therefore unknown to the model. Traditionally, Weibull distribution models were not evaluated based on pre-separated testing data and therefore cannot effectively describe the model's forecasting capability when coping with unknowns. The proposed evaluation method based on testing pairs has the following two advantages:

- A good *MSE* on testing pairs indicate the true forecast accuracy of this model and can describe the forecasting capability for future unknown data using this model. This is especially important for health index based models because a calculated health index using (4) can be any real value in [0,100] and the original asset operating status dataset used for Weibull distribution modelling may not contain these values.
- Furthermore, as explained in Section V, the training pairs are intentionally selected following a uniform distribution so that the corresponding testing pairs will also be able to cover the entire range of available *H* and *F(H)* data. The *MSE* therefore can indicate the forecast capability of the entire failure progression.

Similar ideas of using training and testing sets have been applied very successfully in many supervised machine learning applications [21]. Suitable models are defined as a subset $S' \subset S$ where for a $F_i^s \in S'$, its *MSE* for testing pairs is less than a predefined maximum threshold. The other way to choose $F_i^s$ is to rank all *MSE*s for $F_i, \in S$ and only choose the top models with smallest *MSEs*, for example the top 3 models. Only $S'$ is taken into the next step to establish the joint forecast model.

### B. Combining Weibull Distribution Models as One Joint Model

After evaluating and selecting the suitable Weibull distribution models, this paper further proposes a powerful method to combine different Weibull distribution models as one joint forecasting model. This is a new approach since traditionally only single Weibull distribution model is applied for a forecasting task. One challenge utility asset engineers face is that they cannot pre-determine which Weibull distribution model will be able to yield the most accurate forecast. They also cannot pre-determine the most optimal shifting parameters although they know a probable range these parameters should fall under. A joint model $F_c$, in comparison, can utilize all the suitable models $F_i^s \in S'$ for the final forecast output and is therefore less biased and more robust. Similar ideas can be found in some ensemble learning methods which attempt to build a joint classifier using multiple weaker classifiers [21]. The joint model $F^c$ can be created as:

$$F^c(H) = \sum_{i=1}^{k} F_i^s(H) \left( \frac{\frac{1}{MSE_i}}{\sum_{i=1}^{k} \frac{1}{MSE_i}} \right) \tag{11}$$

In (11), given a new health index input *H*, the output $F_i^s(H)$ of a weaker model with a higher *MSE* will be assigned with a smaller weighting factor while the output $F_i^s(H)$ of a stronger model with a lower *MSE* will be assigned with a larger weighting factor. All suitable models' outputs $F_i^s(H)$ will be averaged by the weighting factors to produce a combined output $F^c(H)$. This $F^c(H)$ is the final forecasted cumulative failure probability for input *H*. In many cases, the joint model has a smaller *MSE* than a single model and is less biased. If $k = 1$, this method is equivalent to selecting the best model for forecasting.

## VII. Asset Reliability Forecast

The joint forecasting model $F^c(H)$ can be used to forecast asset reliability in the future [27].

For a specific future time span $[T_1, T_2]$, a single asset's forecasted failure probability during this time span is:

$$\begin{cases} H_1 = h\ (\Delta T_1, H) \approx H \times \left(\frac{A+\Delta T_1}{A}\right) \\ H_2 = h\ (\Delta T_2,\ H) \approx H \times \left(\frac{A+\Delta T_2}{A}\right) \\ P(H_1,\ H_2) = F^c(H_2) - F^c(H_1) \end{cases} \tag{12}$$

where $\Delta T_1$ and $\Delta T_2$ are the time differences in years from today till $T_1$ and $T_2$; *H* is today's asset health index; $H_1$ and $H_2$ are the future health indexes at $T_1$ and $T_2$; $h(\Delta T, H)$ is a health index estimator and are dependent on the initial health index *H* and time increment $\Delta T$. This estimator can be derived based on statistically checking historical asset health index values that are near *H* and their health variation after a time increment $\Delta T$. It can also be approximated based on the current *H, A* and $\Delta T$ using (12). This approximation is more accurate when $W_a$ is much higher than $W_i$ or when most condition attributes are strongly correlated with age, which in many cases is true.

If only age is used in Weibull distribution modeling, the above equations can be simplified as:

$$\begin{cases} A_1 = A + \Delta T_1 \\ A_2 = A + \Delta T_2 \\ P(A_1,\ A_2) = F^c(A_2) - F^c(A_1) \end{cases} \tag{13}$$

For an asset population with *N* members, the forecasted total number of failures during time span $[T_1, T_2]$ is

$$N_f = \sum_{i=1}^{N} P^i\left(H_1^i,\ H_2^i\right) \tag{14}$$

where $P^i(H_1^i,\ H_2^i)$ is the forecasted failure probability using (12) or (13) for asset member *i,* assuming there are in total *N* members in this asset population.

$N_f$ can be further converted to an asset risk function *R* given the average consequence factor $C_{avg}$ per failure.

$$R = N_f \times C_{avg} \tag{15}$$

TABLE IV
Weighting Factors for Electromechanical Relay's Health Index H

| Item | Weighting factor |
|---|---|
| Age $W_a$ | 0.7 |
| Enclosure Condition $W_1$ | 0.1 |
| Mechanical condition $W_2$ | 0.1 |
| Electrical Condition $W_3$ | 0.1 |

If $C_{avg}$ is unknown or the failure consequence significantly differs among different population members, Monte Carlo simulation can be employed to forecast future system reliability risk for a given time span [28], [29]. This application is further discussed in Section VIII. For a specific asset type, if the numbers of connected customers are known, the total number of customer interruptions can also be forecasted based on $N_f$. This can be further converted to the variation of System Average Interruption Frequency Index (SAIFI). On top of this, if a typical outage duration is assumed, $N_f$ can also be used to calculate the sum of customer interruption durations and the variation of System Average Interruption Duration Index (SAIDI) [4].

## VIII. Case Study

The proposed method was applied to an electric utility company in West Canada for the reliability forecast of electromechanical relays and distribution poles. The process and results are discussed as below.

### A. Electromechanical Relay Reliability Forecast

In this utility company, the electromechanical relay operation dataset contains age data, operating status data and three condition attributes for electromechanical relays: enclosure condition, mechanical condition and electrical condition. The substation technicians annually inspect these electromechanical relays for these three conditions. Based on the inspection results, they assign health ratings for the condition attributes. For a limited population, they also perform a low-voltage simulation test and check if the relay can correctly react to the testing signals and control the trip contacts [30]. If the relay fails to make the correct actions, it will be labeled as failed and will get replaced; otherwise it will be labeled as working. The health index *H* is determined by the normalized relay age, enclosure condition, mechanical condition and electrical condition ratings. Weighting factors are pre-determined by equipment engineers based on domain knowledge and experience. They are listed in Table IV.

The dataset contains 560 relay records with operating statuses obtained from the historical low-voltage simulation tests. The health index results are shown in Fig. 7. Their ages and latest condition ratings are also included. For each relay, the health index *H* is calculated by using (4). Then the dataset can be converted to a cumulative failure probability vs. *H* in the format of Table III. This table is further split into 80 data pairs for training and 20 data pairs for testing.

In this utility company, the electromechanical relays typically have an approximate 2-year energization delay due to substation, transmission and distribution line construction. By checking the inspection records, it is found all relays less than 2-years old have a health index less than 5. Therefore $\gamma$ should be less than 5 and is discretized as $\gamma \in 2.5{,}5$ when choosing two values for study. In addition, the asset engineer is also certain that these electromechanical relays' infant mortality rate is less than 10%, which is discretized as $\delta \in 5\%, 10\%$ when choosing two values for study. Based on these prior knowledge and assumptions, 9 Weibull distribution models are established in the form of (4), (6), (7) and (8). After evaluating each model using the testing pairs, their *MSEs* are calculated and ranked in Table V.

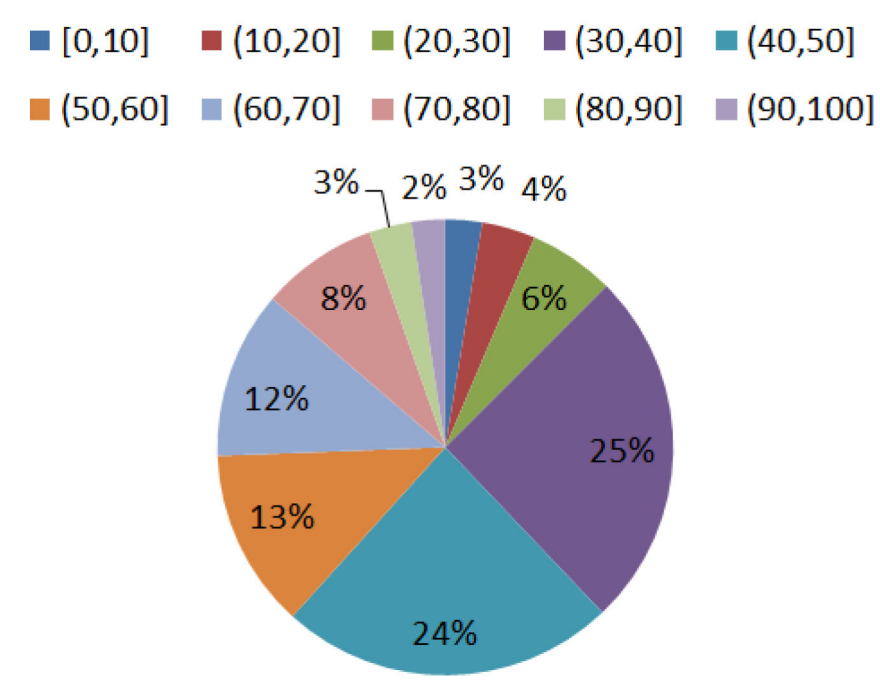

Fig. 7. Health index percentage in the electromechanical relay population.

TABLE V
9 WEIBULL DISTRIBUTION MODELS AND TESTING MSE FOR ELECTROMECHANICAL RELAYS

| ID | Model Description | MSE | Ranking |
|---|---|---|---|
| 1 | Two-parameter | 0.0015 | **3** |
| 2 | X-Shift($\gamma$=2.5) | 0.0015 | 4 |
| 3 | X-Shift($\gamma$=5) | 0.0016 | 6 |
| 4 | Y-shift($\delta$=0.05) | 0.0016 | 5 |
| 5 | Y-shift($\delta$=0.10) | 0.0040 | 9 |
| 6 | XY-shift($\gamma$=2.5,$\delta$=0.05) | 0.0013 | **2** |
| 7 | XY-shift($\gamma$=2.5,$\delta$=0.10) | 0.0035 | 8 |
| 8 | XY-shift($\gamma$=5, $\delta$=0.05) | 0.0013 | **1** |
| 9 | XY-Shift($\delta$=5, $\delta$=0.10) | 0.0034 | 7 |

From Table V, top 3 models are selected as suitable models and are combined as $F^c(H)$ using (11). Graphically, these three suitable models and $F^c(H)$ are shown in Fig. 8. The joint model $F^c$ is also evaluated using the testing pairs and its MSE is 0.0012. This accuracy is better than any single Weibull distribution model listed in Table V.

Accordingly, the utility company could consider preparing at least 277 new relays and replace the failed relays proactively from 2019 to 2023 to minimize the impact of unexpected failures. Some obsolete electromechanical relays may have to be replaced with digital relays. This finding is also important for the utility company to plan spare requirements and construction resources to maintain the desired system reliability.

### B. Distribution Pole Reliability Forecast

The proposed approach was also applied to a sample group consisting of 1000 distribution wood poles selected from one area. Different from electromechanical relays, these poles were installed from recent time to more than 80 years ago and there are no complete inspection records available for producing their health indices. Only installation and replacement records are available. These records indicate the age of poles and the time when some poles were replaced due to failures. In practice, lack of inspection records is a quite common problem for studying aging asset population which covers a relatively long time span. Many old paper based records could be missing. As discussed in Section IV, in cases like this, if the studied population operates in a homogenous environment, age can be used instead of health index to produce modified Weibull distribution models. In this case study, all poles were made of western red cedar, treated with the same type of chemical treatment (Chromated copper arsenate) [31] and carrying straight conductors in one geographic area. Therefore, they can be reviewed as one type of asset and rely on age information for further modeling and forecast.

TABLE VI
5 WEIBULL DISTRIBUTION MODELS AND TESTING MSE FOR DISTRIBUTION POLES

| ID | Model Description | MSE | Ranking |
|---|---|---|---|
| 1 | Two-parameter | 0.0110 | 5 |
| 2 | X-Shift($\gamma$=5) | 0.0093 | **3** |
| 3 | X-Shift($\gamma$=10) | 0.0080 | **1** |
| 4 | X-Shift($\gamma$=15) | 0.0081 | **2** |
| 5 | X-Shift($\gamma$=20) | 0.0101 | 4 |

According to the asset engineer's analysis, these poles do not have infant mortality like the electromechanical relays. The chemical treatment that manufacturers applied to the pole surface could effectively reduce the internal and external infestation for up to 20 years before the treatment gradually degrades under weather exposure and ultimately lose the protection effect. Therefore, excluding rare third-party damages such as vehicle collision, the probability of natural failures of distribution poles are extremely low in an initial period after installation. Based on this important prior knowledge, X-shift parameter $\gamma$ is discretized as $\gamma \in \{5, 10, 15, 20\}$. With the standard two-parameter model, in total 5 Weibull distribution models were established and evaluated. Their *MSEs* are calculated and ranked in Table VI.

From Table VI, top 3 models are selected as suitable models and to produce the joint model $F^c(H)$ as shown in Fig. 9. $F^c$'s MSE is 0.0078 and better than any single Weibull distribution model listed in Table VI.

One application of this joint model is to forecast the need of replacement for distribution poles in other parts of the system that share similar asset properties with the studied sample group. The only data requirement is the age information of the poles. For example, in area B, the demographics of 3000 poles are shown in Fig. 10. Based on the established joint model $F^c(H)$, it is forecasted from 2019 to 2029, in total 263 pole failures could occur. A pole failure often leads to a power outage or public safety compromise. Based on historical incident investigation, it was assumed that the average dollarized consequence per pole failure is \$2,000. Following (15), it can be calculated that the company is expected to incur \$526,000 loss from pole failures in area B from 2019 to 2029. Taking one step further, in order to determine the optimal reliability risk mitigation measures, a Monte Carlo simulation can be conducted along with the established joint probabilistic model $F^c(H)$ to test the effectiveness of these measures. For example, a company could evaluate three proactive pole replacement strategies: replacing 100 oldest poles, 200 oldest poles and 300 oldest poles every year. An average replacement cost can be assumed. Then for each replacement strategy, a Monte Carlo simulation can be applied to simulate the number of pole failures every year after the replacement. Through the comparison of the simulated results, the asset replacement strategy which would result in a minimum overall cost (including failure loss, proactive replacement cost and reactive replacement cost) in the next 10 years can be selected as the optimal replacement strategy.

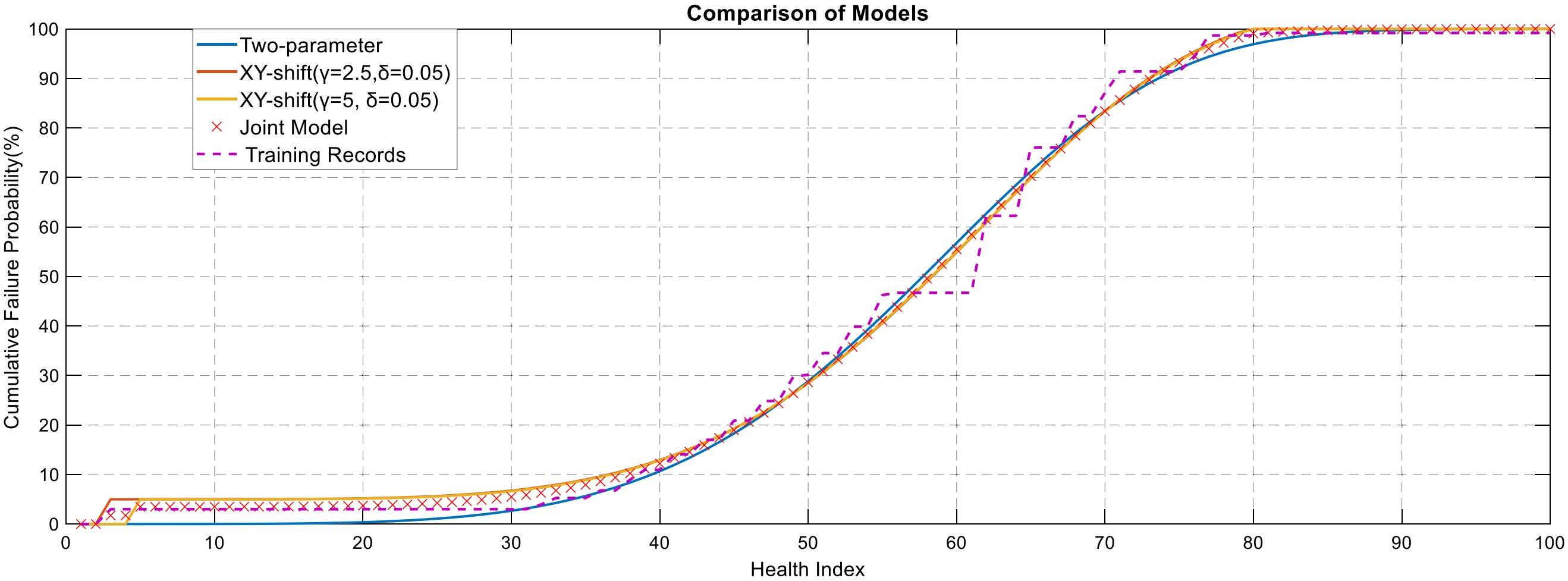

Fig. 8. Comparison of the suitable models and the joint forecast model for Case Study A.

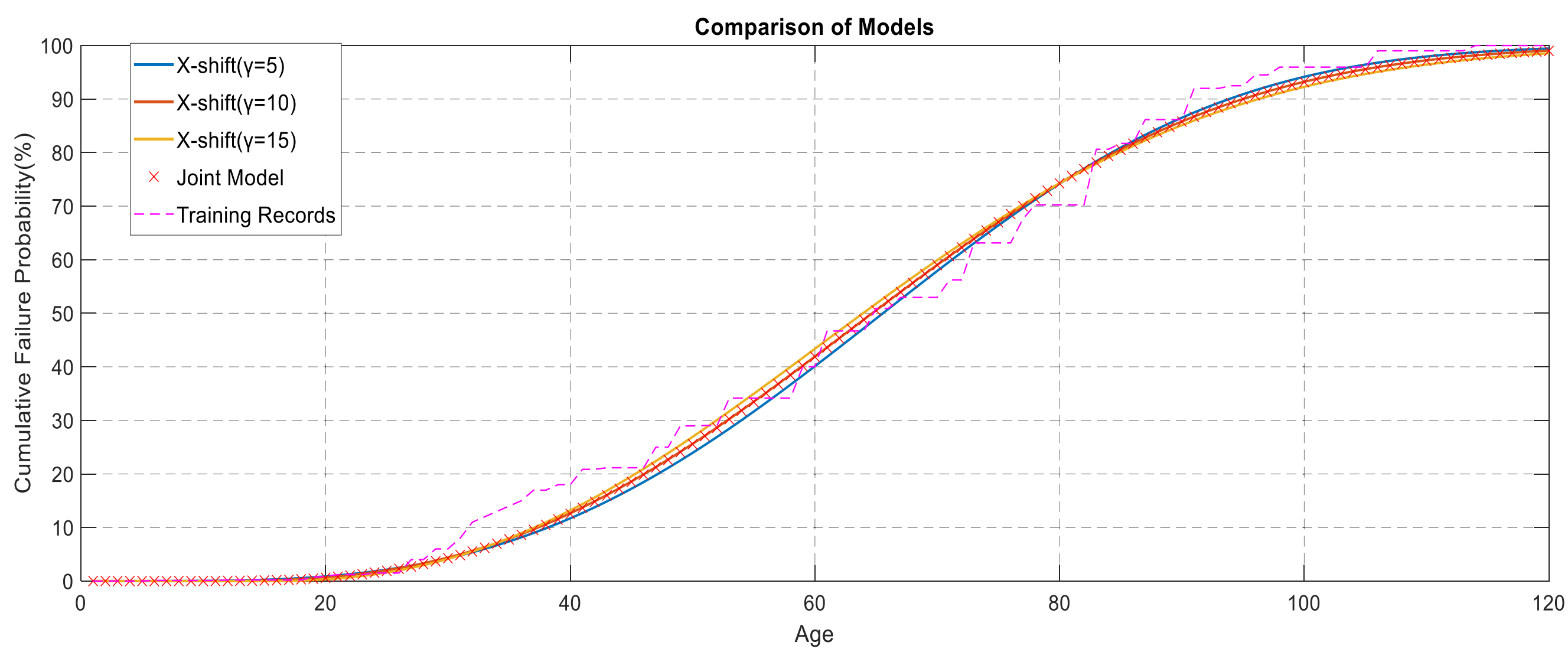

Fig. 9. Comparison of the suitable models and the joint forecast model for Case Study B.

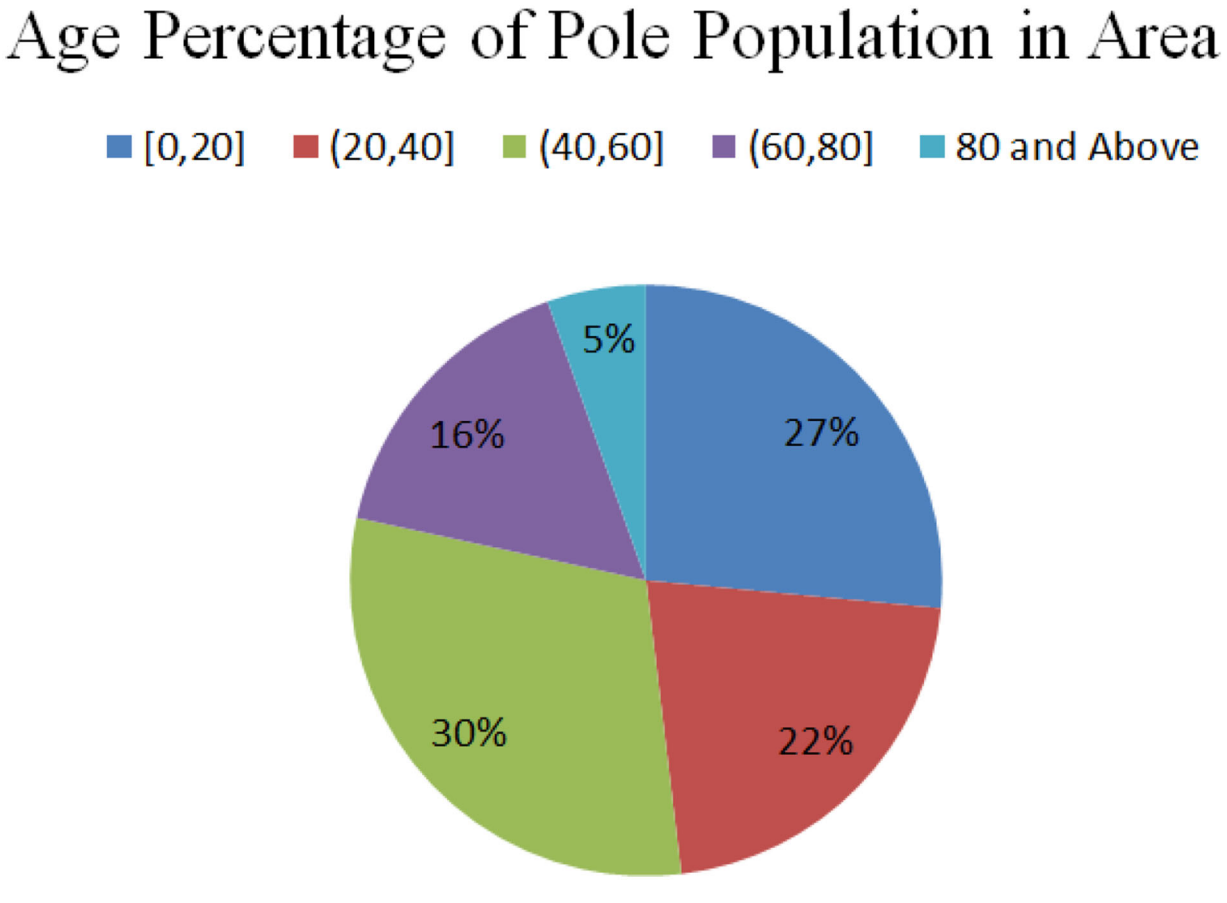

Fig. 10. Age percentage in the distribution pole population in area B.

## IX. Conclusion

This paper presented a novel approach for creating and combining modified Weibull distribution models to forecast power system asset reliability. Compared to previous works, the proposed approach has the following advantages:

- It is not reliant on a single Weibull distribution model. Instead, it can leverage multiple modified Weibull distribution models to incorporate additional information and considerations in the power systems;
- It uses a unique method to measure the performance of different Weibull distribution models. The method can better describe the models' forecasting capability;
- It can effectively combine different Weibull distribution models as a joint forecast model which could have a better performance than individual models.

The proposed approach was applied to a utility company in West Canada to forecast electromechanical relay and distribution pole reliability. Successful results were obtained. The proposed approach can also be used for other power system asset populations as a generic approach.

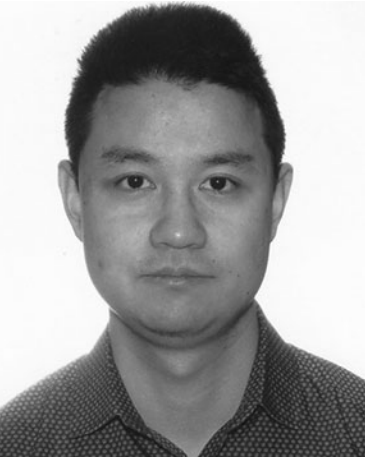

**Ming Dong** (S'08–M'13–SM'18) received the doctoral degree from the University of Alberta, Edmonton, AB, Canada, in 2013. Since graduation, he has been working with various positions in two major electric utility companies in West Canada as a Professional Engineer (P.Eng.) and Senior Engineer for more than 5 years. In 2017, he received the certificate of Data Science and Big Data Analytics from Massachusetts Institute of Technology. He is also a regional officer of Alberta Artificial Intelligence Association. His research interests include applications of artificial intelligence and big data technologies in power system planning and operation, power quality data analytics, power equipment testing, and system grounding.

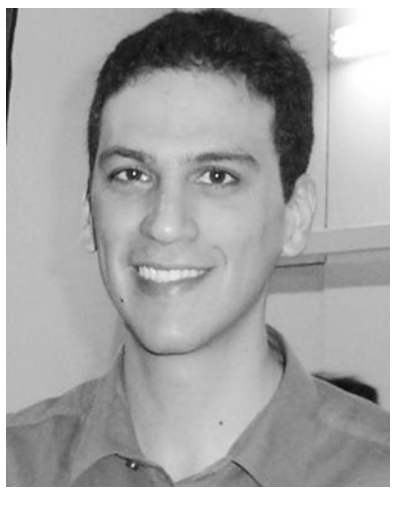

**Alexandre B. Nassif** (S'05–M'09–SM'13) received the doctoral degree from the University of Alberta, Edmonton, AB, Canada and. He is a Specialist Engineer in ATCO Electric. He has authored and coauthored more than 50 technical papers in international journals and conferences in the areas of power quality, DER, microgrids and power system protection and stability. Before joining ATCO, he simultaneously worked for Hydro One as a protection planning engineer and Ryerson University as a Postdoctoral Research Fellow. He is a Professional Engineer in Alberta.